\documentclass[10pt]{article}
\usepackage{amsfonts}
\usepackage{amssymb}
\usepackage{graphicx}
\usepackage{fullpage}
\usepackage{amsmath}
\usepackage{nccmath}
\usepackage{authblk}
\def\be{\begin{equation}}
\def\ee{\end{equation}}
\def\bea{\begin{align}}
\def\eaa{\end{align}}

\def\dd{\partial}

\title{A method to solve nonlinear Schr\"odinger
equation using Riccati equation \footnote{IMSc Preprint Number: IMSC/2014/11/12}} 

\author[1]{Vivek M.
Vyas \thanks{vivekmv@imsc.res.in}}
\affil[1]{Institute of Mathematical Sciences, Taramani, Chennai 600 113, INDIA}
\author[2]{Rama Gupta \thanks{rama10181@davuniversity.org}}
\affil[2]{Department of Physics, DAV University,
Jalandhar 144008, INDIA}
\author[3]{C. N. Kumar \thanks{cnkumar@pu.ac.in}}

\affil[3]{Department of Physics, Panjab University, Chandigarh
160014, INDIA}

\author[4]{Prasanta K. Panigrahi \thanks{pprasanta@iiserkol.ac.in}}
\affil[4]{Indian Institute of Science Education
and Research (IISER) - Kolkata, Mohanpur,
Nadia 741252, INDIA}

\begin{document}

\maketitle

\begin{abstract}
A method to find exact solutions to nonlinear Schr\"odinger equation, defined on a line and on a plane, is found by connecting it with second order linear ordinary differential equation. The connection is essentially made using Riccati equation. Generalisation of several known solutions is found using this method, in case of nonlinear Schr\"odinger equation defined on a line. This method also yields non-singular and singular vortex solutions, when applied to nonlinear Schr\"odinger equation on a plane. 
\end{abstract}

\section{Introduction}

Nonlinear Schr\"odinger equation (NLSE) is one of the well studied
nonlinear partial differential equations (PDE) in the literature. It manifests in a
variety of physical problems, ranging from optical fibres
\cite{agra}, superfluidity \cite{fetter, hua} and ocean waves
\cite{pere} amongst others. When defined on a line, it is found to
be integrable \cite{das}, as was shown by Zakharov and Shabat
\cite{zakh}, using inverse scattering transform. The inverse
scattering transform has yielded several interesting solutions to
this system including the soliton solutions \cite{das}.
This system has also been studied intensely using Hirota's direct
method \cite{hirota} and a number of solutions have been obtained
\cite{hirota2}. Connections of NLSE with second order nonlinear
ordinary differential equations, which are satisfied by elliptic functions, are also well
known and well studied \cite{laksh}. Further, its relation with
Painleve II and IV equations is also known \cite{clark, hieta}.

In this paper, we show that NLSE, defined on a line and on a
plane, in certain cases can be mapped to Riccati equation, a first
order nonlinear ordinary differential equation (ODE). Since Riccati system can be mapped onto a
second order linear ODE, using Cole-Hopf map \cite{pia,khare}, this
effectively gives one a connection between second order linear ODE
and NLSE. Generalised versions of solitonic, periodic and rational
solutions are found for NLSE defined on a line, using this straightforward map. Similar technique is extended to study NLSE defined on a plane, which yields non-singular and singular vortex
solutions.

\section{NLSE on a line}

Nonlinear Schr\"odinger equation, defined on a line, reads
\cite{laksh}: \be \label{nlse} i \frac{\dd \psi}{\dd t} +
\frac{\dd^{2} \psi}{\dd x^{2}} - g|\psi|^2 \psi + \mu \psi = 0,
\ee where both $\mu$ and $g$ are assumed to be positive definite
constants. Two non-dynamical trivial solutions of above equation
are, $\psi = 0$ and $ \psi = \pm \sqrt \frac{\mu}{g}$. Inorder to
find a non-trivial dynamical solution to above equation, we choose
an ansatz: \be\label{ansatz} \psi(x,t) = a \cos{\theta}
+i b \sin{\theta} + d \cos{\theta}
f\left((x-ut)\frac{\cos{\theta}}{\xi}\right ), \ee where $a$, $b$,
$d$, $\xi$ and $\theta$ are real constants, with $f(\tau)$ being a
real function of travelling variable: $\tau
=(x-ut)\frac{\cos{\theta}}{\xi}$. Above ansatz solves equation
(\ref{nlse}) provided following two equations are obeyed by
$f(\tau)$: \begin{align} \label{riccati} & - \frac{ u d {\cos}^{2}
{\theta} }{\xi} f_{\tau} = g b \sin \theta  \left( a^2 \cos^2
{\theta} + b^2 \sin^2 {\theta} + d^2 \cos^2{\theta} f^2 + 2  a d f
\cos^2 {\theta} \right) - \mu b \sin{\theta} \\ \label{jacobi} &
\frac{d \cos^3 \theta }{\xi^2} f_{\tau \tau}  = \left( g (a^2
{\cos}^2 \theta +b^2 {\sin}^2 \theta + d^2 {\cos}^2 \theta f^2 +
2ad \cos^2\theta f) -\mu \right) \left( a \cos\theta + d
\cos\theta f \right). \end{align} Here, $f_{\tau}$ stands for
derivative of $f(\tau)$ with respect to $\tau$. Compatibility of
above two equations requires that $ u = \sqrt{2 g b^2} \sin \theta$.
With this condition, one observes that for all $f(\tau)$ which
solve equation (\ref{riccati}), also solve equation (\ref{jacobi}). 
Hence, by solving for equation (\ref{riccati}), a Riccati equation \cite{pia},
one obtains a corresponding solution to equation (\ref{nlse}). 
Interestingly, this choice of ansatz has mapped the problem
of solving a nonlinear PDE problem given by (\ref{nlse}), to a
problem of solving a Riccati equation.
Riccati equation (\ref{riccati}) can be generally written as,
\be\label{riccatigen} f_\tau = A f^2 + B f + C, \ee where $A
=-\text{sgn}(b)\sqrt{\frac{g}{2}} d \xi$, $B = - \text{sgn}(b) a
\xi \sqrt{2g}$ and $C = - \frac{b |b| \xi \sqrt{2 g}}{d \cos^{2}
\theta} \left( g a^{2} \cos^{2} \theta + g b^{2} \sin^{2} \theta -
\mu \right)$. It is well known that, Riccati equation can be mapped
on to a second order linear ODE via Cole-Hopf map
\cite{pia,khare}: \be \label{colehopf} f(\tau) = -\frac{1}{A}
\frac{u'(\tau)}{u(\tau)}, \ee where $u(\tau)$ solves: \be
\label{sode} u'' - B u' + AC u = 0. \ee

Above is a second order linear ODE, whose general solution is
can be written in terms of two linearly independent solutions,
weighted by two arbitrary constants.

\noindent \underline{\bf{Case I}}: $B^{2} > 4 AC$

In this case, equation (\ref{sode}) is solved by two
linearly independent real solutions $u(\tau) = e^{\lambda_{\pm}
\tau}$, where \be \lambda_{\pm} = \frac{B}{2} \pm
\frac{1}{2}\sqrt{B^{2} - 4 A C}. \ee Hence, the general solution to
equation (\ref{riccatigen}) is given by: \be \label{sol1} f(\tau)
= -\frac{1}{A} \left( \frac{c_{1} \lambda_{+} e^{\lambda_{+} \tau}
+ c_{2} \lambda_{-} e^{\lambda_{-} \tau}}{c_{1} e^{\lambda_{+}
\tau} + c_{2} e^{\lambda_{-} \tau}} \right). \ee Appearance of two
arbitrary constants in above solution may appear unpleasant;
however it is to be noted that, these two will be fixed by appropriate
boundary conditions, under which equation (\ref{nlse}) is solved.
In case when, $a=b=\sqrt \frac{\mu}{g}$, and $c_{1}=c_{2}=1$, one
finds that above solution reduces to the well known grey soliton
solution: \be \psi = a \cos \theta \: \text{tanh} (a A \tau) + i a
\sin \theta. \ee

This shows that the solution (\ref{sol1}) is a generalised version
of known soliton solutions of NLSE \cite{zakh,fadd}.

\noindent \underline{\bf{Case II}}: $B^{2} < 4 AC$

In this case, solution space of equation (\ref{sode}) can be
constructed out of linearly independent solutions $u(\tau) =
e^{\lambda_{\pm} \tau}$, which are in general complex, since \be
\lambda_{\pm} = \frac{B}{2} \pm \frac{i}{2}\sqrt{4 A C - B^{2}}.
\ee The general solution to equation (\ref{riccatigen})
correspondingly is given by: \be \label{sol2} f(\tau) =
-\frac{1}{A} \left( \frac{c_{1} \lambda_{+} e^{\frac{i}{2}\sqrt{4
A C - B^{2}} \tau} + c_{2} \lambda_{-} e^{- \frac{i}{2}\sqrt{4 A C
- B^{2}} \tau}}{c_{1} e^{\frac{i}{2}\sqrt{4 A C - B^{2}} \tau} +
c_{2} e^{- \frac{i}{2}\sqrt{4 A C - B^{2}} \tau}} \right). \ee

Note that, reality of $f$ in this case can be ensured by
appropriate choice of values of constants $c_{1,2}$. As is
evident, this is a generalised periodic solution of NLSE
\cite{fadd}.

\noindent \underline{\bf{Case III}}: $B^{2} = 4 AC$

In this case, equation (\ref{sode}) admits
two linearly independent solutions: $e^{\frac{B}{2} \tau}$ and
$\tau e^{\frac{B}{2} \tau}$. The corresponding solution to
(\ref{riccatigen}) is given by: \be f(\tau) = -\frac{1}{A} \left(
\frac{(c_{1} + c_{2}) + \frac{c_{2} B}{2} \tau }{ c_{1} + c_{2}
\tau} \right). \ee This is a generalised version of known rational
solution of NLSE \cite{hirota2}.

Above one saw, that the ansatz based mapping, defined by (\ref{ansatz}),  from NLSE
problem to a second order linear ODE problem, gives rise to a
variety of generalised solutions to NLSE. Further, it also shows
that solitonic, periodic and rational solutions to NLSE exist in
mutually exclusive parameter space.

\section{NLSE on a plane}

This approach can also be extended to study NLSE defined on a plane. 
Time independent NLSE in two spatial dimensions reads:
\be \label{nlse2}
 - {\nabla}^{2} \psi + g(x, y)|\psi|^2 \psi - \mu(x, y) \psi + V(x, y) \psi = 0,
\ee
where field $\psi$, coupling constant $g$, chemical potential
$\mu$ and trapping potential $V$ are all assumed to be  functions
of coordinates but not of time. In what follows, it will be assumed that one is dealing with
spherically symmetric problems, in which case $g$, $V$ and $\mu$
are all function of $r=\sqrt{x^{2}+y^{2}}$ only. Being interested in
solutions which are isotropic, the following ansatz for $\psi$ is
chosen: \be \label{ansatz2} \psi(r, \theta) = R(r) e^{i n \theta},
\ee where $n$ is an integer, and $R(r)$ is a
real function. In light of this ansatz, equation (\ref{nlse2}) now
reads: \be - \left( \frac{1}{r} \frac{\partial R}{\partial r} +
\frac{\partial^{2} R}{\partial r^{2}} -
\frac{n^{2}}{r^{2}}R\right) + g(r) R^{3} +
\left(V(r)-\mu(r)\right) R = 0. \ee

Above equation can be rewritten in terms of $\rho$, where $R =
r^{-\frac{1}{2}} \rho(r)$, as \be \label{nlse21} \frac{\partial^2
\rho }{\partial r^2} = \frac{g(r)}{r} \rho^3 + \left( \frac{4n^{2}
- 1}{4 r^2} - \mu(r) + V(r) \right)\rho. \ee Often in many
physical problems, one imposes boundary conditions: $\rho(r
\rightarrow \infty) \rightarrow \text{const.}$ and $\rho'(r
\rightarrow \infty) \rightarrow 0$, which shall be assumed to be
true here as well. Equation (\ref{nlse21}) is seen to be equivalent
to following Riccati equation: \be \label{riccati2} \frac{\partial
\rho}{\partial r} = \alpha(r) \rho ^2 + \beta (r) \rho + \gamma
(r), \ee provided following conditions are satisfied:
\begin{align}
&\label{beta-alpha}\beta =-\frac{\alpha'}{3 \alpha}, \\
&\label{beta-gamma} \beta = -\frac{\gamma'}{\gamma},\\
&\label{delta} \alpha^2 = \frac{g(r)}{2 r},\\
&\label{cons-cond.} 2 \alpha \gamma  + \beta' + \beta^2 =  \frac{4n^{2}-1}{4 r^2} - \mu(r) + V(r).
\end{align}
These imply that:
\begin{align}
g(r)&=2 r \alpha^{2}(r),\\
\alpha(r) &= \alpha_{0}\, e^{-3 \int^{r} dt \, \beta(t)},\\
\gamma(r) &= \gamma_{0}\, e^{- \int^{r} dt \, \beta(t)}, \: \text{and}\\
\beta'(r) + \beta^{2}(r) + 2 \alpha_{0} \gamma_{0} e^{-4 \int^{r}
dt \, \beta(t)} & = V(r) + \frac{4n^{2}-1}{4 r^2} - \mu(r).
\end{align} Here, $\alpha_{0},\gamma_{0}$ are constants of
integration and need to be fixed, so as to be compatible with
boundary conditions.

Below two cases are considered, where the above conditions are
explicitly solved, to yield nontrivial solutions to equation
(\ref{nlse2}).

\noindent \underline{\textbf{Case - I}}: $\beta=0$, $\gamma_{0} \neq 0$

When $\beta = 0$, one finds that the above conditions imply, \be 2
\alpha_{0} \gamma_{0} = V(r) - \mu(r) + \frac{4n^{2}-1}{4 r^2},
\ee which is easily satisfied, if the external potential is $V(r)=\frac{1-4n^{2}}{4 r^2}$, 
and the chemical potential is
given by $\mu(r)=-2 \alpha_{0} \gamma_{0}$, where constants
$\alpha_{0}$ and $\beta_{0}$ are nonzero. This also means that
coupling constant $g(r)=2 \alpha_{0}^{2} r$. The Riccati equation
(\ref{riccati2}): \be \frac{\partial \rho}{\partial r} =
\alpha_0 \rho ^2 + \gamma_0, \ee can now be solved by mapping it
to second order linear ODE; \be y'' + \alpha_{0} \gamma_{0} y = 0,
\ee using Cole-Hopf transformation, $\rho = - \frac{1}{\alpha_0}
\frac{y'}{y}$. When $\alpha_{0} \gamma_{0} < 0$, the above linear ODE
admits a solution: \be y(r) = \text{cosh} (\sqrt{|\alpha_{0}
\gamma_{0}|} r). \ee Correspondingly $\rho$ is given by:
\be \rho(r) = -\frac{\sqrt{|\alpha_{0} \gamma_{0}|}}{\alpha_0}
\text{tanh} (\sqrt{|\alpha_{0} \gamma_{0}|} r), \ee
using which $\psi$ comes out to be: \be \psi(r,\theta) =
\frac{-1}{\sqrt{r}} \frac{\sqrt{|\alpha_{0}
\gamma_{0}|}}{\alpha_0} \text{tanh} (\sqrt{|\alpha_{0}
\gamma_{0}|} r) e^{i n \theta}. \ee Note this solution actually
describes a vortex of charge $n$, since gradient of phase
possesses non zero circulation: phase  function $\chi(\theta)=n
\theta$, is such that $\oint \vec{\nabla} \chi \cdot d\vec{l} = 2
\pi n$ along any closed curve enclosing the origin \cite{ume,hua}.
It should be also be noted that density at the core of vortex is zero since
$|\psi(r \rightarrow 0)|^{2} \sim r$, and hence is a non-singular
vortex.

\noindent \underline{\textbf{Case - II}}: $\beta=0$, $\gamma_{0} = 0$

In this case, one finds that
$V(r)=\frac{1-4n^{2}}{4 r^2}$, $\mu = 0$ and $g(r)=2
\alpha_{0}^{2} r$. Riccati equation (\ref{riccati2}) now reads: \be \frac{\partial
\rho}{\partial r} = \alpha_0 \rho ^2. \ee
It is easy to see that, this
equation possesses a power law solution: \be \rho (r) = -
\frac{1}{\alpha_{0} r}, \ee complying with the boundary
conditions. Correspondingly, the expression for $\psi$ reads: \be
\psi(r,\theta) = \frac{-1}{\alpha_{0} r^{3/2} }  e^{i n \theta}.
\ee Akin to the previous case, this solution also describes a vortex of charge
$n$, since here also gradient of phase possesses non-zero
circulation. However unlike above vortex, in this case, density at
the core diverges since $|\psi(r \rightarrow 0)|^{2} \sim
\frac{1}{r^{3}}$. Also note that, unlike earlier case, where a
length scale $\frac{1}{\sqrt{|\alpha_{0} \gamma_{0}|}}$ was
associated with the vortex, often called vortex radius \cite{hua},
here no such scale is present.

This results would have relevance for the discussion regarding various kinds of
possible variation of nonlinearity, like cases of constant nonlinearity and bounded nonlinearity, which 
has been well explored for vortex solutions
\cite{wu2010exact,toikka2012exact}.

\section{Conclusion} In this paper, it is shown that the problem
of solving NLSE, in certain cases, can be mapped onto a problem of
solving a second order linear ODE. The mapping procedure
critically exploits the property of Riccati equation. It is
observed that, for NLSE defined on a line, generalisation of
several known solutions is straightforwardly obtained. In case of
NLSE defined on a plane, two different types of vortex solutions
are found. It must be noted that, a completely different method,
employing the Riccati equation for finding solutions of NLSE type
problem also exists in literature \cite{atre, goyal}. In the
present treatment, we have not dealt with NLSE with a possible
source term \cite{raju1, vyas}. One wonders, if the present
approach can be generalised appropriately to yield solutions to
such inhomogeneous problems as well.


\end{document}